\documentclass[aps,prl,twocolumn,superscriptaddress]{revtex4}
\usepackage{amssymb}
\usepackage{graphicx}
\usepackage{amsmath}
\usepackage{epsf}

\begin{document}

\title{ Quantum phases in a doped Mott insulator on the Shastry-Sutherland
lattice}
\author{ Jun Liu }
\affiliation{Department of Physics and Astronomy and Ames Laboratory, Iowa State
University, Ames, Iowa 50011, USA }
\author{ Nandini Trivedi}
\affiliation{Department of Physics, Ohio State University, Columbus, Ohio 43210, USA}
\author{ Yongbin Lee}
\affiliation{Department of Physics and Astronomy and Ames Laboratory, Iowa State
University, Ames, Iowa 50011, USA }
\author{ B. N. Harmon}
\affiliation{Department of Physics and Astronomy and Ames Laboratory, Iowa State
University, Ames, Iowa 50011, USA }
\author{ J\"{o}rg Schmalian}
\affiliation{Department of Physics and Astronomy and Ames Laboratory, Iowa State
University, Ames, Iowa 50011, USA }
\date{\today}

\begin{abstract}
We propose the projected BCS wave function as the ground state for the doped
Mott insulator \textrm{SrCu}$_{2}\mathrm{{(BO}_{3}{)}_{2}}$ on the
Shastry-Sutherland lattice. At half filling this wave function yields the
exact ground state. Adding mobile charge carriers, we find a strong
asymmetry between electron and hole doping. Upon electron doping an unusual
metal with strong valence bond correlations forms. Hole doped systems are $d$%
-wave RVB superconductors in which superconductivity is strongly enhanced by
the emergence of inhomogeneous plaquette bond order.
\end{abstract}

\maketitle

The physics of doped Mott insulators is at the heart of some of the most
interesting and challenging problems in condensed matter physics\cite%
{Zaanen06}. In the extreme \textquotedblleft atomic" limit, the localization
of electrons in the Mott insulating state leads to a macroscopic degeneracy
of their spin or orbital degrees of freedom. Lifting this degeneracy via
non-local quantum fluctuations gives rise to the rich variety of competing
ground states in real materials. In many cases magnetic or orbital order
prevails. In systems with geometric frustration and strong quantum
fluctuations the possibility of a ground state without such order exists.
Strong quantum mechanical fluctuations are also believed to induce
unconventional superconductivity in doped Mott insulators and are related to
the emergence of resonating valence bond (RVB) type fluctuations\cite%
{Anderson87}. The investigation of doped Mott insulators without long range
magnetic order at half filling is therefore the most direct way to uncover
the nature of the strong local quantum fluctuations and their impact on the
charge carrier dynamics, including superconductivity. It allows one to
address questions such as: \textit{Are all doped RVB states superconductors?}
\textit{What is the role of spatial inhomogeneities?} or \textit{What
determines the stability of states such as the d-wave resonating valence
bond?}. \ By analyzing a specific example, doped \textrm{SrCu}$_{2}\mathrm{{%
(BO}_{3}{)}_{2}}$ with Cu-spins on a Shastry-Sutherland lattice, we give
answers to these interesting questions.

\begin{figure}[tbp]
\includegraphics[width=2.8in]{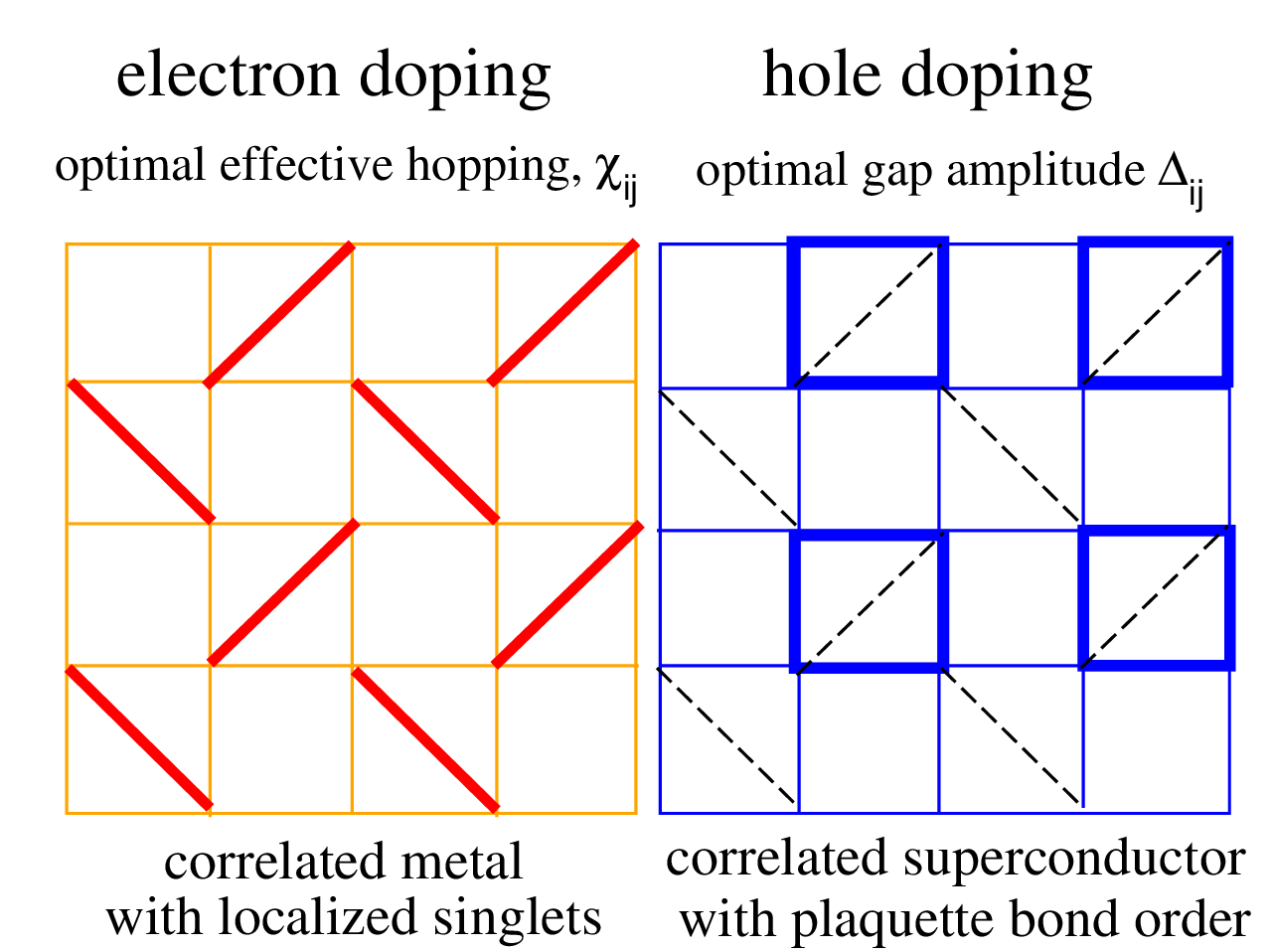}
\caption{Shastry Sutherland lattice with effective hopping $\protect\chi %
_{ij}$ (for electron doping) and local pairing strength $\Delta _{ij}$ (for
hole doping). The width of each line determines the magnitude $\protect\chi %
_{ij}$ or $\Delta _{ij}$, respectively. For electron doping, strong
correlations along the diagonals occur, reminiscent of the undoped valence
bond crystal. For hole doping, the nature of the pairing state changes,
leading to d-wave superconductivity and an inhomogeneous plaquette pattern
of the bond strength. Plaquette states with large $\Delta _{ij}$ around the
two different diagonal orientations are degenerate.}
\label{fig1}
\end{figure}

\textrm{SrCu}$_{2}\mathrm{{(BO}_{3}{)}_{2}}$ is a Mott insulator with a low
temperature spin gap $\Delta _{s}\simeq 30\mathrm{K}$\cite%
{Kageyama99,Miyahara03}. The \textrm{Cu}$^{2+}$ states with $s=\frac{1}{2}$
spins interact predominantly within two dimensional \textrm{CuBO}$_{3}$%
-planes and form dimers along the bond with dominant spin-spin interaction.
The topology of the \textrm{Cu}-sublattice is shown in Fig.~\ref{fig1}. The
dominant, intra-dimer exchange interaction $J^{\prime }$ is along the
diagonal bonds, while the smaller inter-dimer exchange $J$ is along the $x$-
and $y$-axes. There are two remarkable aspects of this material. First, the
exact ground state of the Heisenberg model, on the lattice shown in Fig.~\ref%
{fig1} was obtained by Shastry and Sutherland\cite{Shastry81}. For
sufficiently large $J^{\prime }$ the exact ground state energy per site is $%
E_{0}/N=-\frac{3}{8}J^{\prime }$. The ground state wave function is an
orthogonal dimer state $\left\vert \Phi _{\mathrm{od}}\right\rangle $, i.e.
a direct product of singlets along the bonds connected by $J^{\prime }$: $%
\left\vert \Phi _{\mathrm{od}}\right\rangle =\prod\nolimits_{\left\langle
ij\right\rangle }\left( \left\vert \uparrow \downarrow \right\rangle
_{ij}-\left\vert \downarrow \uparrow \right\rangle _{ij}\right) /\sqrt{2}$.
Shastry and Sutherland demonstrated that this is the ground state for $%
J^{\prime }/J>2.$ More recent numerical results\cite%
{Miyahara99,Weihong99,Koga00,MHartmann00,Weihong02,Lauchli02} support a
transition for $J^{\prime }/J$ between $1.42$ and $1.47$. In the state $%
\left\vert \Phi _{\mathrm{od}}\right\rangle $ singlets are localized, making
this Mott insulator a valence bond crystal without long range magnetic
order. The second interesting aspect of \ \textrm{SrCu}$_{2}\mathrm{{(BO}_{3}%
{)}_{2}}$ is that $J^{\prime }/J\simeq 1.57$ ($J^{\prime }\simeq 85\mathrm{K}
$ and $J\simeq 54\mathrm{K}$\cite{Miyahara03}) is close to the regime where $%
\left\vert \Phi _{\mathrm{od}}\right\rangle $ ceases to be the ground state.
The valence bond crystal is comparatively fragile and competing states
become relevant. Interesting proposals for competing phases at half filling
include helical ordered states\cite{Albrecht96,Chung01}, plaquette singlet
states\cite{Koga00,Chung01,Lauchli02} and a deconfined spinon phase\cite%
{Chung01}.

In this paper we investigate the stability of the valence bond crystal, and
the emergence of superconductivity, flux phases and inhomogeneous states in
doped \textrm{SrCu}$_{2}\mathrm{{(BO}_{3}{)}_{2}}$. We use a resonating
valence bond variational wave function\cite{Anderson87} that includes the 
\emph{exact} ground state energy for the half filled parent compound, making
the approach particularly well controlled for small doping. We demonstrate
that upon adding charge carriers, the system displays a \emph{strong
asymmetry} between electron and hole doping. For electron doping a
non-superconducting but metallic phase with strong valence bond correlations
forms, while an unconventional superconductor emerges for hole doping. In
the latter case a spontaneous breaking of the spatial symmetry occurs, where
the paring gap forms a plaquette pattern, while the charge density remains
uniform (see Fig.~\ref{fig1}). We then show that for hole doping
superconductivity of this inhomogeneous state is \textit{strongly enhanced}
compared to the homogeneous state.

In order to describe doped \textrm{SrCu}$_{2}\mathrm{{(BO}_{3}{)}_{2}}$ we
start from the $t$-$J$ model\cite{Shastry02}: 
\begin{equation}
H=\sum_{ij,\sigma }t_{ij}\widetilde{c}_{i\sigma }^{\dagger }\widetilde{c}%
_{j\sigma }+\sum_{i,j}J_{ij}\left( \mathbf{S}_{i}\cdot \mathbf{S}_{j}-\frac{1%
}{4}n_{i}n_{j}\right) ,  \label{tJ}
\end{equation}%
where $\sum_{\sigma }\widetilde{c}_{i\sigma }^{\dagger }\widetilde{c}%
_{i\sigma }\leq 1$, the tilde indicates that these operate in the projected
Hilbert space, and the remaining notation is standard. In addition to Eq.\ref%
{tJ} we also include three site hopping terms\cite{Paramekanti01} (not shown
explicitly) that are of same order in $t/U$ and become relevant for larger
doping. We include hopping elements $t$ and $t^{\prime }$ for the bonds with
exchange interaction $J$ and $J^{\prime }$, respectively.

Hole doping in the $t$-$J$ model is achieved by taking out electrons. To
describe electron doping we perform a particle-hole transformation and take
out holes. This transformation changes the sign of $t^{\prime }$ but not of $%
t$. Thus, one sign of $t^{\prime }$ corresponds to hole, the other sign to
electron doping. Since $t^{\prime }$ is the largest hopping element we
expect an asymmetry between electron and hole doping. The effect is already
visible if one analyses the single hole state on a $4$-site lattice. A doped
carrier is localized on the diagonals for $t^{\prime }<0$ and delocalizes
for $t^{\prime }>0$. Experiments so far determine $J^{\prime }/J=\left(
t^{\prime }/t\right) ^{2}$ but not the sign of $t^{\prime }$.

\begin{figure}[tbp]
\includegraphics[width=3.in]{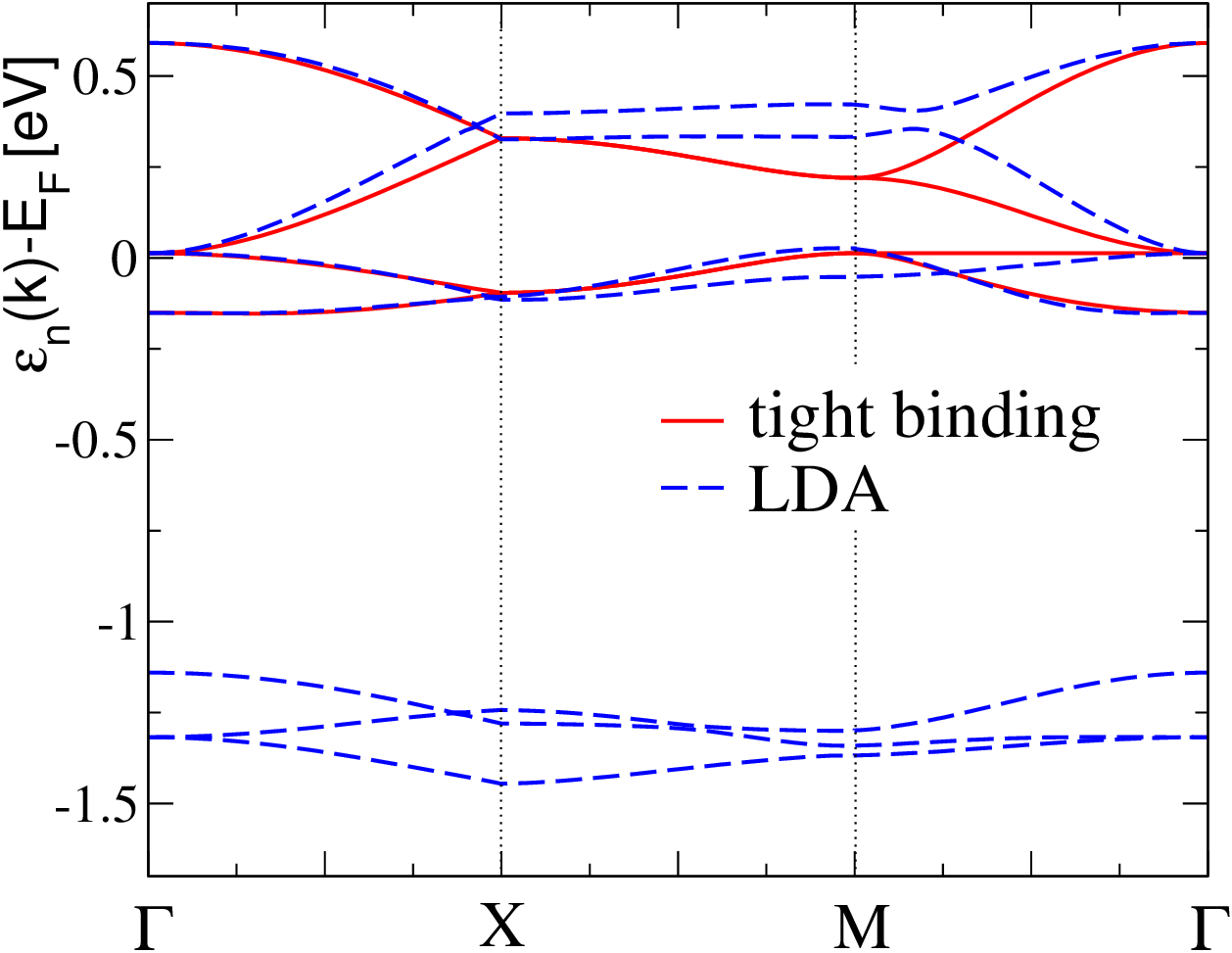}
\caption{Bandstructure of \textrm{SrCu}$_{2}\mathrm{{(BO}_{3}{)}_{2}}$
obtained using density functional theory in comparison with a tight binding
fit. Only the in-plane dispersion of the bands is shown. The bands close to $%
E_{\mathrm{F}}$ are made of $\mathrm{Cu}$-$3d_{x^{2}-y^{2}}$ and $\mathrm{O}$%
-$2p$ states, while the well separated bands $\ $around $-1.25\mathrm{eV}$
consist of $\mathrm{O}$-$2p$ and $\mathrm{Cu}$-$3d_{xz}$ and $3d_{yz}$
states.}
\label{fig2}
\end{figure}

In order to determine the crucial sign \ of the diagonal hopping elements $%
t^{\prime }$ we perform electronic structure calculations for $\ $\textrm{%
SrCu}$_{2}\mathrm{{(BO}_{3}{)}}_{2}$. In cuprate superconductors such
calculations give the proper Fermi surface and band dispersion\cite%
{Pickett89} despite their problems with the Mott insulating state and mass
renormalization effects. We assume the same is true in our case. In Fig.~\ref%
{fig2} we show our results in comparison with a tight binding model where we
ignore correlation effects, making the system a (semi)metal instead of a
Mott insulator. In the LDA calculation we ignore magnetic order. \ Except
for the detailed behavior around the $M$-point the tight binding model gives
a very reasonable description of the electronic structure. It also
demonstrates that other bands, not included in the tight binding approach,
are well separated in energy. A fit close to the $\Gamma $-point yields $%
t=0.09\mathrm{eV}$ and $t^{\prime }=0.104\mathrm{eV}>0$. The ratio $%
t^{\prime }/t$ $\simeq 1.15$ is slightly smaller than the value $1.25$,
obtained from the ratio of the exchange constants. However, a fit where one
forces $t^{\prime }/t=1.25$ still gives a very good description of the LDA
band-structure. In what follows we measure energies in units of $t$, use $%
t^{\prime }=\pm 1.25t$ for hole and electron doping and $J_{ij}=0.3\left%
\vert t_{ij}\right\vert $. The qualitative behavior of our results is
unchanged if we use the experimental values for $J_{ij}$ and the
unrenormalized values for $t_{ij}$ listed above.

We now include the effects of strong correlation into the calculation by
using a resonating valence bond ground state wave function\cite%
{Anderson87,Gros87,Paramekanti01,Anderson04} 
\begin{equation}
\left\vert \Psi \right\rangle =P\left\vert \Phi _{\mathrm{BCS}}\right\rangle
,  \label{rvb}
\end{equation}%
where $P$ projects out all doubly occupied states. $\left\vert \Phi _{%
\mathrm{BCS}}\right\rangle $ is the BCS-ground state wave function. The RVB
description is particularly appealing as it reproduces the exact ground
state at half filling (see below). $\left\vert \Phi _{\mathrm{BCS}%
}\right\rangle $ is the fixed particle projection of the ground state of \ 
\begin{equation}
H_{\mathrm{BCS}}=\sum_{i,j,\sigma }\chi _{ij}c_{i\sigma }^{\dagger
}c_{j\sigma }+\sum_{ij}\left( \Delta _{ij}c_{i\uparrow }^{\dagger
}c_{j\downarrow }^{\dagger }+h.c.\right) .  \label{trial}
\end{equation}%
The effective hopping elements $\chi _{ij}$ and pairing gaps $\Delta _{ij}$
are the variational parameters of our many body wave function and are
determined by minimizing the energy 
$E={\left\langle \Psi \left\vert H\right\vert \Psi \right\rangle }/{%
\left\langle \Psi |\Psi \right\rangle }$. 
$E$ is evaluated using the Monte Carlo approach of Ref.%
\onlinecite{Ceperley77} for $12\times 12$ up to $18\times 18$ lattices. The
many body wave function Eq.\ref{rvb} also allows determination of the long
range correlations between pairs, enabling us to distinguish between local
pairing and off diagonal long range order (ODLRO). Only the latter
corresponds to superconductivity, yields a Meissner effect and flux
quantization. To this end we analyze the pairing correlation function, $F_{%
\mathbf{a,b}}\left( \mathbf{r-r}^{\prime }\right) =\left\langle \Psi
\left\vert b_{\mathbf{r},\mathbf{a}}^{\dagger }b_{\mathbf{r}^{\prime },%
\mathbf{b}}\right\vert \Psi \right\rangle $, with $b_{\mathbf{r},\mathbf{a}}=%
\frac{1}{2}\left( c_{\mathbf{r+a,\downarrow }}c_{\mathbf{r\uparrow }}-c_{%
\mathbf{r+a,\uparrow }}c_{\mathbf{r,\downarrow }}\right) $ and $\mathbf{a}$, 
$\mathbf{b}$ nearest neighbor vectors. $\left\vert \psi _{s}\right\vert
^{2}\propto F_{\mathbf{a,b}}\left( \mathbf{R={r-r}^{\prime }\rightarrow
\infty }\right) $ determines the superconducting order parameter $\psi _{s}$.

\emph{Undoped system}: At half filling, the variational wave function in Eq.%
\ref{rvb} yields the exactly known ground state energy $E_{0}$ with an
accuracy $10^{-7}$. We further obtain the expected spin correlations $%
\left\langle \mathbf{S}_{i}\cdot \mathbf{S}_{j}\right\rangle $ which equals $%
-\frac{3}{4}$ for sites $i$ and $j$ on the same dimer bond and vanishes
otherwise with error bars of $10^{-4}$. Thus, at half filling the RVB wave
function, Eq.\ref{rvb}, reproduces the exact ground state of the
Shastry-Sutherland model, making it an ideal starting point to investigate
doped systems. 
\begin{figure}[tbp]
\includegraphics[width=2.8in]{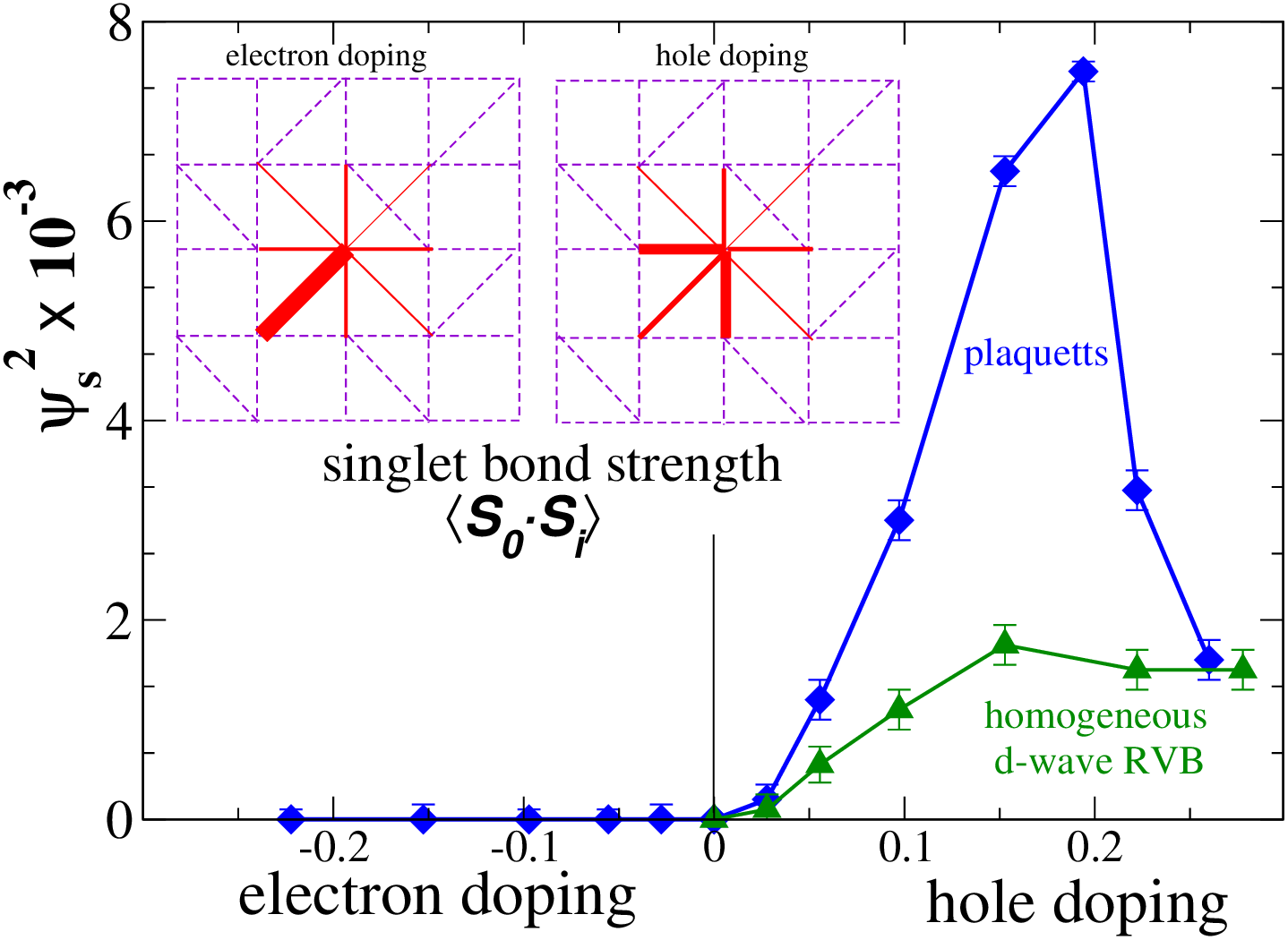}
\caption{Superconducting order parameter, signaling ODLR for hole doped
systems. Even though the wave function for electron doped systems is
characterizd by a pairing strength $\Delta _{ij}\neq 0$ in the BCS wave
function, no superconducting order exists. From the analysis of the low
energy Drude weight we conclude that the system is a metal. For hole doped
systems $d$-wave superconductivity emerges. The superconducting order
parameter is strongly enhanced in the inhomogeneous plaquette phase. The
inset shows the local spin correlations for electron and hole doping. The
thickness of the line refers to the magnitude of $\langle \mathbf{S}_{0}%
\mathbf{S}_{i}\rangle $. All nearest and next nearest neighbor correlations
are antiferromagnetic. }
\label{fig3}
\end{figure}

\emph{Doped valence bond crystal}: At finite electron doping, we find that
the renormalized diagonal hopping $\chi ^{\prime }$ that enters $\left\vert
\Phi _{\mathrm{BCS}}\right\rangle $ as a variational parameter is
significantly enhanced compared to the bare values. For doping values $%
x\simeq 0.1$ we find $\chi ^{\prime }/\chi \simeq 1.5t^{\prime }/t$, see
also the left panel of Fig.~\ref{fig1}. The spin correlations $\left\langle 
\mathbf{S}_{i}\cdot \mathbf{S}_{j}\right\rangle $ shown in the inset of Fig.~%
\ref{fig3}, are large for sites $i$ and $j$ on the same diagonal bond and
very small otherwise and mimic the behavior of the orthorgonal dimer state
of the undoped parent compound.

The pairing gap $\Delta _{ij}$ is finite (for example, at electron doping $%
x=0.1$, we obtain the gaps along diagonal bonds $\Delta _{i,i+x\pm y}\approx
0.3t$ ). It is remarkable that this many body state with finite local
pairing gap does not show any indication for off diagonal long range order
(see Fig.~\ref{fig3}). The pairing correlation function $F_{\mathbf{a,b}}(%
\mathbf{{R})}$ rapidly vanishes beyond the nearest neighbor. We also
determined the low energy contribution of the Drude weight\cite%
{Paramekanti01} and found $D_{\mathrm{Drude}}\simeq 3.18xt$ for small $x$.
Despite the addition of mobile charge carriers, singlets form on diagonal
bonds and the system remains in a non-superconducting but metallic phase
with strong valence bond correlations. The absence of superconductivity for $%
t^{\prime }>0$ is consistent with the finding of Ref.\cite{Leung04}.

\emph{Correlated Superconductor}: The hole doped case, on the other hand,
shows superconductivity (see Fig.3). The pairing gap has $d_{x^{2}-y^{2}}$
symmetry with $\Delta _{i,i+\mathbf{x}}=-\Delta _{i,i+\mathbf{y}}$. Also its
magnitude is larger for hole doping compared to the electron doped case (we
find $\Delta _{i,i+\mathbf{x}}\simeq t$ for hole doping $x=0.1$). The
magnetic correlations in this superconductor are very different from the
valence bond crystal. As shown in the inset of Fig.~\ref{fig3}, we find $%
\left\langle \mathbf{S}_{i}\cdot \mathbf{S}_{j}\right\rangle <0$ with rather
comparable magnitude for \emph{all} neighbors, even when they are coupled
via the exchange interaction $J$.

As pointed out above, the origin for the strong asymmetry between hole and
electron doping is due to the constructive \ ($t^{\prime }>0$) and
destructive ($t^{\prime }<0$) interference of hopping paths in the system.
This leads to a delocalization of singlets and a true RVB state for hole
doping while for electron doping the underlying valence bond crystal is only
moderately affected.

\emph{Inhomogeneous Superconductor}: So far we assumed that the system is
fully homogeneous, i.e. that no spatial symmetry in the problem is broken.
Next we allow for the local potentials $\chi _{ii}$, effective hopping
elements $\chi _{ij}$ as well as pairing gaps $\Delta _{ij}$ to vary in
space. Specifically, we allow for arbitrary values of these quantities
inside the crystallographic unit cell, consisting of four \textrm{Cu}-sites.
For electron doping no additional symmetry is broken. In case of hole
doping, we find that the spontaneous emergence of inhomogeneous solutions
leads to a reduction of the ground state energy ($\Delta E\simeq 4.3\times
10^{-3}t$ for $x=0.1$ which is three times the energy gain of the
homogeneous $d$-wave state compared to the $\Delta _{ij}=0$ state). While
the charge is homogeneous, we find large variations of the pairing gaps
following the pattern indicated in the right panel of Fig.\ref{fig1}. For $%
x\simeq 0.1$ we find for strong bonds a gap amplitude $\Delta \simeq 2.3t$
compared to $\simeq 0.2t$ for weak bonds. The plaquette state with periodic
variation of the pairing strength is driven by a gain in the kinetic energy.

In this plaquette state, the effective hybridizations $\chi ^{\prime }$ on
the diagonals also varies in space. $\chi ^{\prime }$ on diagonals
surrounded by large pairing gaps is reduced compared to $\chi ^{\prime }$ on
plaquettes with small gaps. At half filling, an analysis of the
Shastry-Sutherland lattice with \textrm{Sp}$\left( N\right) $ symmetry\cite%
{Chung01} led to several plaquette states, including one similar to ours,
once strong quantum fluctuations were taken into account. Doping with charge
carriers is clearly one route to enhance quantum fluctuations. Physically, $%
d $-wave RVB fluctuations between the states $\left\vert =\right\rangle $
and $\left\vert \shortparallel \right\rangle $ inside a plaquette seem
strongest if the system is allowed to break up in a spatially inhomogeneous
pattern.

The most interesting aspect of this new plaquette ordered state is that long
range superconducting order $\psi _{s}$ is strongly enhanced in the
plaquette state compared to the homogeneous state as seen in Fig.~\ref{fig3}%
. The pairing correlation function $F_{\mathbf{a,b}}\left( \mathbf{R}\right) 
$ reaches its long distant limit after several lattice constants. Strong
pairing in plaquettes separated by less that this superconducting coherence
length is therefore very efficient to boost a superconducting state\cite%
{Martin95}. For hole doping the magnitude $\left\vert \psi _{s}\right\vert
^{2}$ is approximately $10\%$ of the value one obtains within BCS-theory for
the same pairing gap. For electron doping the effect of quantum fluctuations
much more dramatic since $\left\vert \psi _{s}\right\vert ^{2}=0$ even
though a BCS theory with finite gap would always yield a finite order
parameter.

A simple analytic approach that provides useful insight in case of other
doped Mott insulators is the slave boson mean field theory. The approach is
closely related to the wave function, Eq.\ref{rvb}, with the key difference
that the projection $P$ is done only on the average. Performing a slave
boson calculation, we find, in agreement with earlier work\cite{Shastry02},
that the strong asymmetry between electron and hole doping does not emerge.
The plaquette state found in this paper does not follow from the slave boson
calculation either. Finally, by using the full projection we searched
unsuccessfully for a staggered flux phase with $\chi _{ij}=\chi e^{\pm
i\theta }$ , found in a slave bosons calculation\cite{Chung04}. These
results demonstrate that proper projection, which enters the slave boson
theory via strong gauge field fluctuations, is crucial for the
Shastry-Sutherland lattice.

In summary, we have identified a doped Mott insulator with a very rich
behavior as function of varying charge carrier concentration. Electron doped
systems display metallic phases with strong valence bond correlations and
are the first example for a doped Mott insulator where an RVB-wave function
yields local pairing but no superconductivity. Hole doped systems enter a
d-wave RVB state, become superconducting and an inhomogeneous bond order
emerges spontaneously which enhances superconductivity. From the energy gain
of the paired state and the values of the exchange energies $J_{ij}$ we
estimate a superconducting transition temperature $T_{c}$ of several Kelvin.
Experimentally, attempts to dope \textrm{SrCu}$_{2}\mathrm{{(BO}_{3}{)}_{2}}$
have not led to metallic behavior\cite{GTLiu05}. However, recent advances in
the sample preparation of transition metal oxides\cite{Takagi02} give us
every reason to be optimistic that the obstacles to dope this material can
be overcome. We hope that our results stimulate research in this direction.

We are grateful to David C. Johnston who initially encouraged us to
investigate this problem. In addition we are grateful to I. Affleck, C.
Batista and M. Randeria for helpful discussions. This research was supported
by the Ames Laboratory, operated for the U.S. Department of Energy by Iowa
State University under Contract No. DE-AC02-07CH11358 and by a Fellowship
from the Institute for Complex Adaptive Matter (JL). We acknowledge the use
of computational facilities at the Ames Laboratory.

\end{document}